\documentclass[12pt]{article}
\usepackage{epsf,graphicx,multirow}
\usepackage{epsfig}
\newcommand{\beq}{\begin{equation}}
\newcommand{\eeq}{\end{equation}}
\newcommand{\ba}{\begin{array}{c}}
\newcommand{\ea}{\end{array}}
\newcommand{\be}{\begin{eqnarray}}
\newcommand{\ee}{\end{eqnarray}}

\newcommand{\br}{\mbox{$^{8}{B}~$}}
\newcommand{\ber}{\mbox{$^{7}{Be}$~}}

\newcommand{\kl}{\mbox{KamLAND~}}

\newcommand{\ms}{\Delta m^2_{21}}

\newcommand{\sss}{\sin^2 \theta_{12}}

\newcommand{\dms}{\Delta m_{21}^2}
\begin{document}
\begin{flushright}
SISSA 44/2006/EP     \\
TIFR/TH/06-23
\\ 
TUM-HEP-646/06 \\
\end{flushright}

\vspace{0.4cm}
\begin{center}
{\Large \bf
Solar Model Parameters and Direct Measurements of Solar
Neutrino Fluxes}

\vspace{0.5cm}

Abhijit Bandyopadhyay$^{a)}$,
Sandhya Choubey$^{b,c)}$,
Srubabati Goswami$^{b,d)}$, 
S. T. Petcov$^{e)}$
\footnote{Also at: Institute of Nuclear Research and
Nuclear Energy, Bulgarian Academy of Sciences, 1784 Sofia, Bulgaria.} \\

\vskip 0.5cm

$^{a)}${\it Tata Institute of Fundamental Research, 
Mumbai 400005, India}
\\
$^{b)}$ {\it  Harish-Chandra Research Institute, 
Allahabad 211 019, India.} 
\\
$^c${\it Theoretical Physics,}
{\it University of Oxford,}
{\it 1 Keble Road, Oxford OX1 3NP, UK.}
\\
$^{d)}${\it Physik-Department T30d,
Technische Universitaet Muenchen, 
D-85748 Garching, Germany. }
\\
$^{e)}${\it Scuola Internazionale Superiore di Studi Avanzati and\\
Instituto Nazionale di Fisica Nucleare, I-34014 Trieste, Italy.}

\vskip 0.8cm
\end{center}
\begin{abstract}
We explore a novel possibility of 
determining the solar model parameters, 
which serve as input in the calculations 
of the solar neutrino 
fluxes, by exploiting the data from direct  
measurements of the fluxes.
More specifically, we use the rather 
precise value of the 
$^8B$ neutrino flux, $\phi_B$ 
obtained from the global 
analysis of the solar neutrino and KamLAND data,
to derive constraints on each of 
the solar model parameters on which 
$\phi_B$ depends.
We also use more precise values of $^7Be$ and $pp$ 
fluxes as can be
obtained from future prospective data
and discuss 
whether such measurements can help in reducing
the uncertainties of one or more  
input parameters of the Standard Solar Model. 

\end{abstract}
\newpage
\section{Introduction}

 There has been a remarkable  
progress in the studies of solar neutrinos
in the last several years.
The evidences of solar neutrino 
($\nu_{\odot}$) oscillations,
obtained first in the Homestake 
experiment and strengthened by 
the results of Kamiokande, 
SAGE and GALLEX/GNO collaborations
\cite{sol}, were made subsequently compelling 
by the data of  Super-Kamiokande (SK), 
SNO and KamLAND (KL) experiments
\cite{SKsol02,SNO123,KL162}
\footnote{We would like to recall 
that the hypothesis of neutrino 
oscillations was formulated in \cite{BPont57}.
In \cite{BPont67} it was suggested that 
the solar $\nu_e$ 
can take part in 
oscillations involving another active
or sterile neutrino. The article 
\cite{BPont67} appeared
before the first data of the
Homestake experiment were reported.
For a detailed discussion of the
evolution of the solar neutrino problem
since 1970 and of the variety of different 
``solutions'' proposed over the years, 
see, e.g., \cite{BiPet87,STPSchlad97}.}.
The combined charged current (CC) and 
neutral current (NC) data from
SNO and the $\nu$-$e^-$ elastic scattering data 
from SK experiment 
showed that the solar $\nu_e$ undergo 
flavour conversion on their way from 
the central part of the Sun, where they 
are produced, to the Earth.
Under the plausible 
assumption of CPT-invariance, 
the results of the KL reactor
neutrino experiment \cite{KL162} 
established the large mixing 
angle (LMA) MSW 
oscillations/transitions 
as the dominant mechanism
at the origin of the observed 
solar $\nu_e$ deficit.
The existing global neutrino oscillation 
data allow us to conclude that
the solar $\nu_e$ undergo 
transitions (predominantly) 
into almost an equal 
mixture of $\nu_{\mu}$ and $\nu_{\tau}$
neutrinos.
The ratio of the CC and NC event rates observed 
at SNO provided a measure of the 
solar $\nu_e$ transition probability
at energies of $E\sim (5 - 10)$ MeV,
while the SNO NC data permitted to
determine with rather good precision 
the $^8$B component of the solar $\nu_e$ flux
\cite{SNO4}.
The global solar neutrino data allowed to 
obtain information on the other important components 
of the solar neutrino flux - the fluxes of  
$pp$, $pep$ and $^7$Be neutrinos, and to 
constrain the flux of $CNO$ neutrinos 
\cite{roadmap}.
The combined solar neutrino and KamLAND data 
lead to a determination of the 
neutrino oscillation parameters
which drive the solar $\nu_e$ oscillations -
the neutrino mass squared difference $\Delta m^2_{21}$ and
the mixing angle $\theta_{12}$, with an 
unexpectedly high precision (see, e.g., \cite{Bandyopadhyay:2004da, others}).

   The latest SNO result on the $^8$B neutrino flux, $\phi_{B}$,
as reported in \cite{SNO4}, reads
\begin{eqnarray}
\phi_{B}^{NC} &=& 4.94 (1\pm0.088) \times 10^{6} ~{\rm cm^{-2}s^{-1}}\;.
\label{phinc}
\end{eqnarray}
%
This is in good agreement with the
Standard Solar Model (SSM)  prediction 
\cite{bp04}
\begin{eqnarray}
\phi_{B}^{SSM} &=&
5.79 (1\pm 0.23) \times 10^{6} ~{\rm cm^{-2}s^{-1}}\;.
\label{phi8bssm}
\end{eqnarray}
%
The global oscillation analysis of 
the solar neutrino and \kl data performed in 
\cite{Bandyopadhyay:2004da}, 
in which  $\phi_{B}$
is treated as a free parameter, 
yields the following 
value of the flux: 
\begin{eqnarray}
\phi_{B}^{Global} &=& 4.88 
(1\pm0.036) \times 10^{6}~ {\rm cm^{-2}s^{-1}}\;.
\label{phiglobal}
\end{eqnarray}
%
The value of $\phi_{B}$ thus obtained 
corresponds to $f_B = 0.84 $, where 
$f_B$ is defined as 
\begin{eqnarray}
f_B = \frac{\phi_B}{\phi_B^{SSM}}\;.
\label{fb}
\end{eqnarray}
%
   The values of the  $pp$ and $^7$Be neutrino fluxes 
can also be determined similarly  
from a global analysis of the solar neutrino 
and \kl data, in which
the solar luminosity constraint is used 
\cite{roadmap}.
This analysis showed that with the current data sets 
the pp neutrino flux can be determined with an 
uncertainty  
which is the same as 
the estimated uncertainty in the 
SSM prediction for the flux \cite{bp04}.
The uncertainty obtained for the \ber flux   
in the same analysis
is larger than the estimated one in the 
SSM prediction for the same flux \cite{bp04}.
Future high precision measurements of the 
$^7$Be neutrino flux by Borexino and \kl can 
lead to a reduction of the
uncertainties in both \ber and $pp$ fluxes
determined from the solar neutrino data.

    In the present article we explore the 
possibility of using the
precision data (current and prospective) on the 
i) $^8$B, ii) $^8$B and $^7$Be, and 
iii) $^8$B, $^7$Be and $pp$, solar neutrino fluxes 
in order to obtain ``direct'' information 
(i.e., to constrain or determine) on at least 
some of the solar model parameters -  nuclear reaction rates,
opacity, diffusion, heavy element surface abundance, 
etc., which enter into the calculations of  
the fluxes in the Standard Solar Model (SSM).
It is important to establish whether 
high precision measurements of the 
$^8$B, $^7$Be and $pp$ neutrino fluxes
can provide also significant constraints on
the indicated SSM parameters because,
none of the latter can be determined directly experimentally.
The relevant nuclear reaction rates
are measured in direct experiments but at energies
which are considerably higher than the energies 
at which the reactions proceed in the central 
part of the Sun. As a consequence, one has to 
employ an extrapolation procedure (based on 
nuclear theory) in order to
obtain the values of the rates at
the energy of interest.
The solar luminosity $L_{\odot}$
is measured directly with very high accuracy. 
However, it is still important 
to determine this fundamental solar
observable from solar neutrino 
flux measurements. The latter provide
``real time'' information on the rates of 
the nuclear fusion reactions in 
the central region of the Sun. Both the 
photons, observed in the form of 
solar luminosity, and the neutrinos, 
emitted by the Sun, are simultaneously 
produced in these reactions.
It takes neutrinos approximately 8 minutes 
to reach the Earth. In contrast, the ``conventionally'' 
measured luminosity of the Sun
is determined by photons emitted by
the solar surface, 
which, however, were produced in the central 
region of the Sun 
$\sim 4\times 10^4$ years earlier - 
the time it takes these photons to reach the 
surface of the Sun (see, e.g., \cite{Bahcall:2004mz}).
Thus, a comparison of the experimentally measured 
solar luminosity with that obtained from
neutrino flux measurements allows, in particular,
to test the thermo-nuclear fusion 
theory of energy generation
in the Sun and the hypothesis that the Sun,
in what concerns the energy generation,
taking place in its interior, 
and the energy radiation from its surface, 
is in an approximate steady state.

 Another SSM parameter on which the solar 
neutrino fluxes depend is 
the ratio of the surface abundance in mass of 
the elements heavier than helium 
and of the surface abundance (in mass) of hydrogen
(surface heavy element composition), $Z/X$. 
At present there is rather large uncertainty
in the SSM estimated value of this parameter
(see, e.g., \cite{bp04,BSeren04}), 
as will be discussed in somewhat greater detail 
further. Moreover, the estimated uncertainty in 
the value of  $Z/X$ 
is obtained \cite{bp04,BSeren04}~
assuming that the total spread of 
all modern determinations of $Z/X$ 
is equal to the 3$\sigma$ uncertainty in $Z/X$. 
Clearly, a determination of the surface element
composition parameter $Z/X$ from neutrino flux 
measurements could be very helpful for resolving 
the indicated problems. It could also be 
very useful for solar model building. 

   In the analysis that follows we will use the
SSM by Bahcall and Pinsonneault from 2004 \cite{bp04}
as a ``benchmark'' solar model for the
predictions of the solar neutrino fluxes and the 
estimated uncertainties in these predictions, 
originating from the different SSM parameters.

%
\section{Preliminary Observations}
%

   There are 
six principal nuclear reactions and decays in which
neutrinos are produced in the Sun (see, e.g., \cite{BP92}). 
Four of them
generate neutrinos with continuous energy spectrum. These are 
the fusion of two protons ($pp$ $\nu$'s), 
and the decays of the nuclei $^8B$ ($^8B$ $\nu$'s), 
$^{13}N$ and $^{15}O$ ($CNO$ $\nu$'s).
The other two, the fusion of two 
protons and electron 
and the capture of an electron by a $^7Be$ nucleus 
produce neutrino lines 
(the so-called $pep$ and $^7Be$ $\nu$'s). 
The shapes of the energy spectra of 
the $pp$, $^8B$ and
$CNO$ neutrinos
are determined by nuclear physics and are well known. 
However, the SSM predictions for 
the total values of the 
$pp$, $pep$, $^7Be$,
$^8B$ and the $CNO$ 
neutrino fluxes depend 
on several SSM input parameters.  
The uncertainties associated with these
parameters lead to (normalisation) uncertainties 
in the predicted fluxes.  

 There are altogether 11 input SSM parameters 
on which the SSM predictions for the 
$^8B$, $^7Be$, $pp$ and the $CNO$
solar neutrino fluxes in general 
depend  \cite{bp04}. These are first of all
the S-factors (see, e.g., \cite{BBP01}) of the nuclear reactions
$^1H(p,e^+\nu_e)^2H$, $^3He(^3He,2p)^4He$,$^3He(^4He,\gamma)^7Be$,
$^{14}N(p,\gamma)^{15}O$, $^7Be(p,\gamma)^8B$,
and of $e^-$ capture on  $^7Be$.
They are standardly denoted respectively 
by $S_{11}$, $S_{33}$, $S_{34}$, $S_{1,14}$, $S_{17}$
and $S_{e^-7}$. 
The additional parameters
are directly related to the physics of 
the Sun: they are \cite{bp04} the solar luminosity, $L_{\odot}$, 
age, $\tau_{\odot}$, opacity, $O_{\odot}$,
diffusion, $D_{\odot}$, and the ratio of the mass fractions of 
the elements heavier than helium and of hydrogen
at the surface of the Sun (surface composition), $Z/X$ . 
The type of dependence of a given solar neutrino flux
($^8B$, $^7Be$, $pp$, $pep$ and $CNO$)
on a specific SSM parameter varies with the flux.
For the three fluxes of interest for 
our further discussion, for instance, 
we have \cite{BBP01}: 
\beq
\ba
\phi_{B} = C_B~
(S_{11})^{-2.59} 
(S_{33})^{-0.40}
(S_{34})^{+0.81}
(S_{1,14})^{+0.01} 
(S_{17})^{+1.0}
(S_{e^-7})^{-1.0}  \\ [0.3cm]
\times (L_{\odot})^{+6.76}
(\tau_{\odot})^{+1.28}
(O_{\odot})^{-2.93}
(D_{\odot})^{-2.20}
(Z/X)^{+1.36}\;,
\label{PhiB}
\ea
\eeq
\beq
\label{PhiBe}
\phi_{Be} = C_{Be}~
(S_{11})^{-0.97} 
(S_{33})^{-0.43}
(S_{34})^{+0.86}
(L_{\odot})^{+3.40}
(\tau_{\odot})^{+0.69}
(O_{\odot})^{-1.49}
(D_{\odot})^{-0.96}
(Z/X)^{+0.62}\;,
\eeq
\beq
\phi_{pp} = C_{pp}~
(S_{11})^{+0.14} 
(S_{33})^{+0.03}
(S_{34})^{-0.06}
(S_{1,14})^{-0.02} 
(L_{\odot})^{+0.73}
(\tau_{\odot})^{-0.07}
(O_{\odot})^{+0.14}
(D_{\odot})^{+0.13}
(Z/X)^{-0.08}\;,
\label{Phipp}
\eeq
%
where $C_B$, $C_{Be}$ and $C_{pp}$ are constants.

 The nuclear reaction $S$-factors of interest 
are measured typically at c.m. energies 
exceeding $\sim 500$ keV, which are at least by a factor 
$\sim 25$ larger than those corresponding to the conditions 
under which the reactions take place in the central region 
of the Sun. As a consequence, the results on the $S$-factors
 obtained experimentally have to be extrapolated to the much
lower energy of $\sim 20$ keV, which is of interest for the 
solar neutrino flux calculations. The extrapolation procedure
brings additional theoretical uncertainty in the $S$-factor
values. This uncertainty can be substantial for a given 
$S$-factor.

  Because of the importance of the $S_{17}$-factor 
for the prediction of $\phi_{B}$ and the interpretation 
of the data of the Homestake, SK and SNO experiments,
considerable efforts have been made
to determine it with a relatively high precision. 
This was done using data i) from a direct measurement
of the cross-section of the 
reaction $p + ^7Be \rightarrow ^8B + \gamma$ \cite{junghans}, 
and ii) of indirect studies of the same reaction
(via the Coulomb dissociation process 
$\gamma + ^8B  \rightarrow p + ^7Be$ \cite{coulomb}, and
heavy-ion transfer and breakup processes \cite{heavy,azhari}).
In the SSM calculations of the solar neutrino fluxes \cite{bp04} 
the most precise result reported in  
\cite{junghans} is used. 
The value recommended in \cite{junghans} reads:
\be 
S_{17}(0) = 21.4 \pm 0.5~(expt)~\pm 0.6~(theo)~{\rm eV~b},
\label{s17recent}
\ee
%
the quoted 1$\sigma$ error being smaller than 5\% 
\footnote{The earlier recommended value \cite{adelberger},
$S_{17}(0) = 19^{+4}_{-2}$ eV b, had 
a 1$\sigma$ uncertainty of approximately 15\%.}.
However, from a more recent data, obtained with radioactive 
ion beams, the following value was found in \cite{azhari}: 
$S_{17}(0) = 18.2 \pm 1.7$ eV b. 
One of the theoretical uncertainties in 
the value of $S_{17}$ quoted above,
eq. (\ref{s17recent}), is associated with the 
extrapolation method used to obtain this result. 
In \cite{gai} it is argued that a larger extrapolation 
error, than is usually taken into account,
should be assigned in the evaluation 
of the uncertainties in $S_{17}(0)$. 

  According to ref. \cite{bp04}, 
the largest contribution to the 
uncertainties in the predictions of the fluxes 
of neutrinos produced in the reactions of the  
$pp$-chain in the Sun ($^8B$, $^7Be$, $pp$, $pep$), 
is due to the uncertainty in the knowledge  
of the S-factor $S_{34}$.
The estimated  1$\sigma$ 
uncertainty in $S_{34}$ is approximately 9.4\% \cite{bp04}.
The $S_{34}$ factor has been measured recently
\cite{He3He4Singh04} at four values of the
c.m. energy from the interval
(400 - 950) keV. Using an extrapolation
of earlier and the new results
to an energy of $\sim 20$ keV,
the authors of \cite{He3He4Singh04} obtain
$S_{34} = 0.53 \pm 0.02 \pm 0.01$.
If this result will be confirmed,
that would lead to a reduction of the 
1$\sigma$ uncertainty in the value of $S_{34}$
approximately by a factor of 2.
The LUNA collaboration
is planning to perform a more precise
measurement of $S_{34}$ in an experiment 
which is under preparation at the Gran-Sasso 
National Laboratory in Italy \cite{luna}. 

 In the case of the CNO neutrinos,
the largest nuclear physics uncertainty is 
associated with the uncertainty in the value 
of the cross-section of the reaction 
$p + ^{14}{N} \rightarrow ^{15}{O} + \gamma$,
which is directly related to the S-factor $S_{1,14}$
(see Table 2). 

  In what concerns the uncertainties 
in the predictions of the solar neutrino fluxes 
due to the other non-nuclear physics parameters,
the largest, according to the SSM BP04 \cite{bp04},
is a consequence of the lack of sufficiently 
precise knowledge of the surface element 
composition of the Sun.
New values for the abundances in mass of the elements 
$C$, $N$, $O$, $Ne$ and $Ar$ have been derived 
\cite{Asplund05}
using three-dimensional rather than one-dimensional
model of the solar atmosphere, including hydrodynamical 
effects, etc. The new abundance estimates together
with the best-estimates for other solar abundances 
\cite{GrevS98} imply $Z/X = 0.0176$, which is
considerably smaller than the earlier result 
\cite{GrevS98} $Z/X = 0.0229$. 
The new values of the solar surface 
abundances of $C$, $N$, $O$, $Ne$ and $Ar$,
when incorporated into solar models,
lead to serious discrepancies with
helioseismological data 
\cite{bp04,bss,BSeren04}~
\footnote{For this reason the authors of 
our ``benchmark'' SSM 
BP04 did not include the 
new most recent estimate of 
the parameter $Z/X$ in the calculations 
of the solar neutrino fluxes.
The possible effects of this new
result on the SSM predictions of the solar 
neutrino fluxes were considered in \cite{bss,BSeren04}.    
}.
The estimated uncertainty in the value of 
$Z/X$, according to \cite{bp04}, 
is approximately 15\% (see Table 2). 
The latter is obtained \cite{bp04,BSeren04}~
\footnote{The authors of \cite{BSeren04} made a
comment worth quoting concerning the uncertainties 
in the element abundances under discussion:
``Estimating the uncertainty in an abundance 
determination is even more difficult than 
arriving at a best-estimate abundance.''.}
assuming that the total spread of 
all modern determinations of $Z/X$ 
is equal to the 3$\sigma$ uncertainty in $Z/X$. 

 It is pertinent to point out that
the knowledge of the solar neutrino oscillation 
parameters $\Delta m^2_{21}$ and $\sin^2\theta_{12}$
with relatively small uncertainties 
can be crucial for a successful 
high precision determination of the fluxes of 
solar $^8B$, $^7Be$ and $pp$ neutrinos.
The existing data allow a determination of 
$\Delta m^2_{21}$ and $\sin^2\theta_{12}$
at 3$\sigma$ with an  error of 
approximately 11\% and 25\%, respectively. 
Much higher precision can (and most likely will)
be achieved in the future.
The data from phase-III of 
the SNO experiment~\cite{SNO4} using 
$^3$He proportional counters
for the neutral current rate measurement
could lead to a reduction of the error 
in $\sin^2\theta_{12}$ 
to 21\%~\cite{SKGdCP04,BCGPTH1204}. 
If instead of 766.3~t~yr one uses simulated 3~kt~yr
KamLAND data in the same global solar and reactor neutrino 
data analysis, the 3$\sigma$ errors in 
$\Delta m^2_{21}$ and $\sin^2\theta_{12}$ 
diminish to 7\% and 18\%~\cite{BCGPTH1204}. 
The most precise measurement of 
$\Delta m^2_{21}$,
could be achieved \cite{SKGdCP04} using 
Super-Kamiokande doped with 0.1\% of Gadolinium 
(SK-Gd) for detection of reactor 
$\bar{\nu}_e$ \cite{SKGdBV04}:
the SK detector gets the same flux of reactor
$\bar{\nu}_e$ as KamLAND and
after 3 years of 
data-taking, $\Delta m^2_{21}$
could be determined with  
an error of 3.5\% at 3$\sigma$ 
\cite{SKGdCP04}. 
A dedicated reactor  
$\bar{\nu}_e$ experiment with a 
baseline $L\sim 60$~km, tuned to the minimum of the
$\bar{\nu}_e$ survival probability, 
could provide the most precise determination of $\sin^2\theta_{12}$
\cite{TH12}: with statistics of $\sim 60$ GW~kt~yr and a systematic error
of 2\% (5\%), $\sin^2\theta_{12}$ could be measured with an accuracy of
6\% (9\%) at 3$\sigma$ \cite{BCGPTH1204}.  The inclusion of the
uncertainty in $\theta_{13}$ ($\sin^2\theta_{13}<$0.05) in the
analyzes increases the quoted errors by (1--3)\% to approximately 9\%
(12\%) \cite{BCGPTH1204}. 
Even higher precision in the measurement of 
$\Delta m^2_{21}$ and $\sin^2\theta_{12}$ can be reached
with one module (of 147~kt fiducial mass)
of the water \v{C}erenkov detector MEMPHYS 
doped with 0.1\% of Gadolinium (MEMPHYS-Gd), 
and with a 50~kt scale liquid scintillator detector (LENA), 
installed in the Frejus underground laboratory \cite{MEMGd06}. 
The improved determination of $\dms$ and
$\theta_{12}$ with KamLAND or dedicated post-KamLAND reactor neutrino
experiments has been studied also, e.g., in
refs.~\cite{Bandyopadhyay:2003ks,Minakata:2004jt,Kopp:2006mw},
whereas the potential improvements of the precision on these
parameters from future solar neutrino experiments has been
investigated, e.g., in refs.~\cite{BCGPTH1204,TH12,roadmap}.

%
\section{Present Knowledge and Future 
Measurements of Solar Neutrino Fluxes} 
%
%
  Precise knowledge of solar neutrino fluxes  
is a key ingredient of our analysis. 
In this Section,  we summarise the presently 
existing data and information on solar neutrino fluxes 
and the improvements 
in the determination of the fluxes 
that are possible in the future. 

\begin{figure}[t]
\begin{center}
\epsfig{file=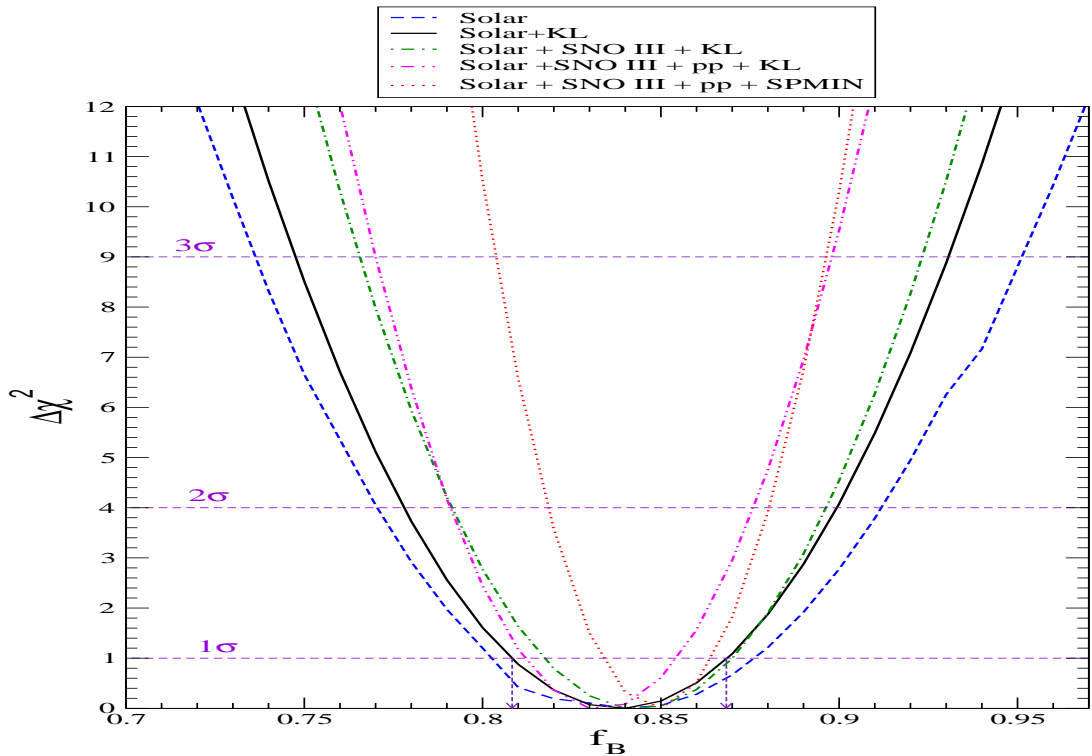,width=6in,height=4in}
\caption{\label{fig:fb}
The dependence of $\Delta \chi^2$ on $f_B$, showing 
the range of allowed values of $f_B$, 
determined using the currently existing data 
and prospective data from future experiments. 
}  
\end{center}
\end{figure}

  Among all the eight solar neutrino fluxes, at present 
we only have a direct experimental determination of the total 
\br neutrino flux produced inside the Sun, through the 
measurement of the rate of the
neutral current reaction on deuterium in the SNO experiment 
(cf. eq. (\ref{phinc})). 
The  $1\sigma$ uncertainty in the value of the $^8B$ neutrino 
flux determined from the NC SNO data 
is approximately 8.8\% (see eq. (\ref{phinc})). 
Global oscillation analyses of solar and  
\kl data, in which the \br neutrino flux
is treated  as a free parameter,
allow to determine $\phi_B$
with even higher precision,
as eq. (\ref{phiglobal}) shows.
In Fig. \ref{fig:fb} we present 
the range of allowed values of 
$f_B$ (defined in eq. (\ref{fb})),
obtained in analyzes of the current data 
and of prospective data from future experiments. 
The corresponding $1\sigma$ 
uncertainties in $f_B$ are given in Table \ref{tab:fb}.
Combining the results from other solar neutrino experiments 
with the NC data from SNO improves 
the precision of determination of the solar 
neutrino oscillations parameters
$\ms$ and $\sss$ and hence reduces the 1$\sigma$ 
error on the $^8B$ neutrino flux to 4.4\%. 
This is further reduced to 3.6\% by 
addition of the \kl results in the global data analysis.
Also shown in Fig. \ref{fig:fb} and Table \ref{tab:fb} are
the expected uncertainty on $f_B$ with inclusion of 
prospective data from future planned/proposed experiments. 
The phase-III of SNO is expected to provide 
direct (uncorrelated) 
measurement of the NC event rate 
with a precision greater than that achieved in 
the earlier salt phase. For the third phase data from SNO
we have assumed the same central values of the 
CC and NC event rates as those observed during the salt phase,
but smaller uncertainties in the measured CC and NC rates,
namely, 4.0\% and 6.4\% 
respectively \cite{poon}. 
With the inclusion of the indicated prospective 
results from this phase 
(referred to as SNO-III in Fig. \ref{fig:fb} 
and Table \ref{tab:fb}),
the $1\sigma$ uncertainty in $f_B$ 
could be reduced to 3.2\%. 
This uncertainty would further 
diminish to 2.5\% and 1.7\% if we 
added successively to the analysis 
the prospective data from a ``generic $pp$''
and from ``SPMIN'' experiments. 
The ``generic $pp$'' experiment  
in Fig. \ref{fig:fb} and Table \ref{tab:fb}
refers to a high precision 
$\nu - e^-$ elastic scattering experiment 
\footnote{There are a number of planned 
sub-MeV solar neutrino experiments
(LowNu Experiments) 
aiming to observe and measure directly 
the $pp$ neutrino flux
using either charged current reactions
(LENS, MOON, SIREN \cite{lownu})
or the $\nu - e^-$ elastic scattering process
(XMASS, CLEAN, HERON, MUNU, GENIUS \cite{lownu}).
} 
with  assumed 1\% error in the measured reaction rate.
The latter is simulated at the best-fit values of
$\ms=8\times 10^{-5}$ eV$^2$ and $\sss=0.31$. 
The ``SPMIN'' refers to a reactor experiment 
with a baseline of 60 km, tuned
to the Survival Probability MINimum \cite{TH12}. 
The results given in Fig. \ref{fig:fb} 
and Table \ref{tab:fb}
for this experiment correspond to 
statistics of 3 kt yr of data  
and a systematic error of 2\%. 
We refer the reader to 
\cite{BCGPTH1204,TH12} for 
further details of our method 
used for the analysis of projected 
data from future experiments.

\begin{table}
\begin{center}
\begin{tabular}{|c|c|}
\hline
Data set used & 1$\sigma$ uncertainty in $f_B$ (in \%)\\
\hline
Solar & 4.4 \\
Solar + KamLAND & 3.6  \\
Solar + SNO-III + KamLAND & 3.2 \\
Solar + SNO-III + pp + KamLAND  & 2.5\\
Solar + SNO-III + pp + SPMIN & 1.7 \\
\hline
\end{tabular}
\end{center}
\caption{\label{tab:fb} 1 $\sigma$ uncertainties of $f_B$}
\end{table} 
%
The SSM prediction for \ber neutrino flux reads:
\be
\phi_{Be}^{SSM}
&=&
4.86(1\pm 0.12) \times 10^{9} ~{\rm cm^{-2}sec^{-1}}~,
\label{phi7bessm}
\end{eqnarray}
%
the estimated uncertainty being 12\%. 
The flux of low energy $pp$ neutrinos 
is calculated very precisely within the SSM - the
estimated uncertainty is 1\%:
\begin{eqnarray}
\phi_{pp}^{SSM}
&=&
5.94 (1\pm 0.01)~\times 10^{10} {\rm cm^{-2}sec^{-1}}~.
\label{phippssm}
\end{eqnarray}
%

 The \ber and $pp$ neutrino fluxes can be determined 
in a solar model independent way together with the 
\br neutrino flux
by treating all three fluxes as free parameters  
in global oscillation analyzes \cite{roadmap}. 
Below we summarise the results obtained for the
\ber and $pp$ neutrino fluxes  
using present and future prospective data from 
solar neutrino experiments \cite{roadmap}. 
The results for $\phi_{pp}$ depend on whether the luminosity 
constraint \cite{lum} is included in the 
analysis or not; the determination
of $\phi_{Be}$ is essentially independent   
of the luminosity constraint \cite{roadmap}. 
Without employing the luminosity constraint,  
both $\phi_{Be}$ and $\phi_{pp}$ are 
determined from the existing data with a precision
which is much worse than the estimated precision of the
BP04 SSM predictions for the two fluxes.
The inclusion of the
luminosity constraint in the analysis
brings about a drastic improvement 
in the determination of $\phi_{pp}$: 
even with the present 
solar and \kl data, $\phi_{pp}$ is determined
with an uncertainty of approximately 
$2\%$ at 1$\sigma$. 

  Improvement in the determination of $\phi_{Be}$
will be possible from a direct measurement of the 
\ber neutrino flux, as is envisaged in the forthcoming 
Borexino solar neutrino \cite{borexnu2004} and $\kl$ 
experiments \cite{klnu2006}.
Data from the Borexino experiment 
with a total 1$\sigma$ error of 10\%
could lead to a determination of the 
$^7Be$ neutrino flux with a $1\sigma$ 
uncertainty of $\sim 10\%$, which is somewhat 
smaller than the estimated uncertainty 
in the BP04 SSM prediction for $\phi_{Be}$ 
(see eq.(\ref{phi7bessm})).
With a total uncertainty of 5\% in the measured event 
rate in Borexino, $\phi_{Be}$ is expected to be 
determined with 1$\sigma$ error of 5.5\%,
while a measurement of the 
Borexino event rate with an error of 3\% 
could lead to a factor 
of three improvement in the precision  
on $\phi_{Be}$ with respect to the currently 
estimated 12\% uncertainty in the SSM 
prediction for $\phi_{Be}$. 

  Precision data from the Borexino experiment
is expected to bring also a
significant improvement in the precision 
of determination of $\phi_{pp}$. As long 
as the luminosity constraint is imposed, 
a measurement of the Borexino event rate 
with a 5\% error could 
lead to a determination of $\phi_{pp}$ 
with a 1$\sigma$ uncertainty of $0.5$\%.
The addition of a high precision 
data from a ``generic $pp$'' experiment
is not expected to lead to any significant 
reduction of the uncertainty in the value of 
$\phi_{pp}$ as long as the luminosity constraint 
is taken into account \cite{roadmap}.
However, the data from these experiments
could certainly improve noticeably 
the precision of the $pp$ flux determination, 
which can be achieved without using 
the luminosity constraint.
Moreover, it should be possible 
to test the applicability of the 
photon luminosity constraint itself
and, more generally, the thermo-nuclear 
theory of the energy generation in the Sun, 
by comparing the measured value 
of the solar photon luminosity 
with the value obtained using
the results from the LowNu and 
the other solar neutrino 
experiments \cite{roadmap}. 

\section{Determining 
the SSM Input Parameters 
from Direct Solar Neutrino Flux Measurements}
\label{nparam}

 In the theoretical framework of 
the SSM, the dependence of
the solar neutrino flux from
the $i^{th}$ nuclear process, $\phi_{i}$,
on the input parameters of the SSM is given,
as we have already seen on the examples of 
the $^8B$, $^7Be$ and $pp$ fluxes, by power laws:
\begin{eqnarray}
\phi_i &=&  C_i \times
\prod_{{\rm all~}j} x_{j}^{\alpha_{ij}}\;.
\label{flux}
\end{eqnarray}
%
Here $C_i$ is a constant   
and $\alpha_{ij}$ in the exponent is the 
logarithmic derivative of
$\phi_i$ with respect to the SSM input parameter $x_j$,
\begin{eqnarray}
\alpha_{ij} &=& \frac{\partial \ln \phi_i}{\partial \ln x_j} .
\end{eqnarray}
%
The values of the different logarithmic derivatives
for the BP04 SSM \cite{bp04} are given
in Table \ref{tab_alpha}.  
We have, in general, eight such equations 
for the eight different solar neutrino fluxes
\footnote{In this counting we have included also the
fluxes of $hep$ and $^{17}F$ solar neutrinos \cite{bp04},
which, however, are predicted to be exceedingly small.
These two fluxes do not play any significant 
role in the analyses of the
presently existing solar neutrino data.}. 

  Let us first consider a general formulation of the problem where 
we assume that we have a total of N input parameters 
in the solar model calculations and  
that out of the total eight fluxes, K  different fluxes
along with their 1$\sigma$
uncertainties can be determined 
from direct measurements. 
We can then pick up the set of K  equations from
the above set (\ref{flux}) 
and can solve them for any subset
$x_{j_1},x_{j_2},..,x_{j_K}$ of 
K different input parameters of the solar model. 
For the rest of (N-K) parameters  
${x_j}$  we can use the values ${x_j}^{0}$ 
found in the SSM. 
We can express the constants $C_i$'s in terms
$\phi_i^{SSM}$'s and $x_j^0$'s using the equation
\begin{eqnarray}
\phi_i^{SSM} &=& C_i \times  
\prod_{{\rm all~}j} (x_j^0)^{\alpha_{ij}}~.
\label{fluxssm}
\end{eqnarray}
%
Thus, we write the set of K equations for the fluxes 
$\phi_{i_1},\phi_{i_2},...,\phi_{i_K}$ as
\begin{eqnarray}
\frac{\phi_i}{\phi_i^{SSM}}
&=&
\frac
{{\prod \atop {j=j_1,j_2,..,j_K}}(x_j)^{\alpha_{ij}}
{\prod \atop {j\neq j_1,j_2,..,j_K}}(x_j^0)^{\alpha_{ij}}}
{{\prod \atop {{\rm all~}j}} (x_j^0)^{\alpha_{ij}}} \nonumber \\
&=&
\prod_{j= j_1,j_2,..,j_K}
\left(\frac{x_j}{x_j^0}\right)^{\alpha_{ij}},
\qquad i=i_1,i_2,..,i_K\;.
\label{phi_i}
\end{eqnarray}
%
Taking logarithm on both sides of eq.\ (\ref{phi_i}) we get
\begin{eqnarray}
\ln\left(\frac{\phi_i}{\phi_i^{SSM}}\right)
&=&
\sum_{j=j_1,j_2,..,j_K}\alpha_{ij} \ln\left(\frac{x_j}{x_j^0}\right),
\qquad i=i_1,i_2,..,i_K\;.
\label{absvalueeqn1}
\end{eqnarray}
%
The above set of K equations can be written in a matrix form as
\begin{eqnarray}
\pmatrix{
\alpha_{i_1j_1} & \alpha_{i_1j_2} &  .. & \alpha_{i_1j_K} \cr
\alpha_{i_2j_1} & \alpha_{i_2j_2} &  .. & \alpha_{i_2j_K} \cr
 .  & . & . & . \cr
 . & . & . & . \cr
\alpha_{i_Kj_1} & \alpha_{i_Kj_2} &  .. & \alpha_{i_Kj_K} 
}
\pmatrix{
\ln \frac{x_1}{x_1^0} \cr
\ln \frac{x_2}{x_2^0} \cr
. \cr
. \cr
\ln \frac{x_K}{x_K^0} 
}
&=&
\pmatrix{
\ln \frac{\phi_1}{\phi_1^{SSM}} \cr
\ln \frac{\phi_2}{\phi_2^{SSM}} \cr
. \cr
. \cr
\ln \frac{\phi_K}{\phi_K^{SSM}} 
}
\label{phimat}
\end{eqnarray}
%
A general solution of  eq.\ (\ref{phimat}) is given by
\begin{eqnarray}
\ln \left(\frac {x_r}{x_r^0}\right)
&=&
\frac{Det X_r}{Det A}, \qquad r=j_1,j_2,..,j_K\;,~~~{\rm Det} A \neq 0\;,
\label{gensolphi}
\end{eqnarray}
%
where
\begin{eqnarray}
A &=&
\pmatrix{
\alpha_{i_1j_1} & \alpha_{i_1j_2} & .. & \alpha_{i_1j_K}\cr
\alpha_{i_2j_1} & \alpha_{i_2j_2} & .. & \alpha_{i_2j_K}\cr
 . & . & . & . \cr
 . & . & . & . \cr
\alpha_{i_Kj_1} & \alpha_{i_Kj_2} & .. & \alpha_{i_Kj_K}
}
\label{amat}
\end{eqnarray}
%
and $X_r$ is the matrix obtained by replacing the $r^{th}$ column of
the matrix $A$ by the column matrix in the right hand side of 
eq. (\ref{phimat}). Thus, we can write the solution 
for the set of $K$ parameters 
of the solar model $\{x_r\}$, $r=j_1,j_2,..,j_K$, in the form
\begin{eqnarray}
x_r &=& x_r^0 \exp\left(\frac{Det X_r}{Det A}\right), \qquad
r=j_1,j_2,..,j_K\;.  
\label{xgensol}
\end{eqnarray}
%

 To evaluate the uncertainties in each of the K SSM parameters of the
set $\{x_j\}$, $j=j_1,j_2,..,j_K$, determined using the data on the
solar neutrino fluxes, we take the derivative of
the logarithm of both sides of eq.\ (\ref{flux}) for $i=i_1,i_2,..,i_K$:
\begin{eqnarray}
\delta\ln\phi_i
&=& \sum_{{\rm all~}j} \alpha_{ij} \delta \ln x_j\;,
\qquad i=i_1,i_2,..,i_K\;.
\label{error(phi_i}
\end{eqnarray}
%
This set of K equations can also be written in matrix form as
%
\begin{eqnarray}
\pmatrix{
\alpha_{i_1j_1} & \alpha_{i_1j_2} & .. & \alpha_{i_1j_K}\cr
\alpha_{i_2j_1} & \alpha_{i_2j_2} & .. & \alpha_{i_2j_K}\cr
 . & . & . & . \cr
 . & . & . & . \cr
\alpha_{i_Kj_1} & \alpha_{i_Kj_2} & .. & \alpha_{i_Kj_K}
}
\pmatrix{
\delta \ln x_{j_1} \cr
\delta \ln x_{j_2} \cr
. \cr
. \cr
\delta \ln x_{j_K} \cr
} 
&=&
\pmatrix
{
\delta \ln \phi_{i_1} 
- {\sum\atop{j\neq{j_1,.., j_K}}}\alpha_{i_1j}\delta \ln x_j \cr
\delta \ln \phi_{i_2}
- {\sum\atop{j\neq{j_1,.., j_K}}}\alpha_{i_2j}\delta \ln x_j \cr
.\cr
.\cr
\delta \ln \phi_{i_K} 
- {\sum\atop{j\neq{j_1,.., j_K}}}\alpha_{i_Kj}\delta \ln x_j \cr
}
\label{errormat}
\end{eqnarray}
%
A general solution of  eq.\ (\ref{errormat}) is then given by
\begin{eqnarray}
\delta \ln x_r &=& \frac{{\rm Det} D_r}{{\rm Det} A}\;, \qquad
r =j_1,j_2,..,j_K\;,~~~{\rm Det} A\neq 0\;,
\label{dxgensol}
\end{eqnarray}
%
where $A$ is the matrix given by eq. (\ref{amat}) and $D_r$ is the
matrix obtained by replacing the $r^{th}$ column of the matrix $A$ by
the column matrix appearing in the right hand side of 
eq. (\ref{errormat}). Obviously,
the right hand side of eq. (\ref{dxgensol})
should be a linear combination of the quantities 
$\frac{\delta \phi_i}{\phi_i}$, $i=i_1,i_2,..,i_K$, and 
$\frac{\delta x_j}{x_j}$, $j\neq j_1,j_2,..,j_K$, and 
therefore eq. (\ref{dxgensol}) can be written as
\begin{eqnarray}
\delta \ln x_r &=&
\sum_{i=i_1,..,i_K} P_{ir} \delta \ln \phi_i
+ \sum_{j\neq j_1,..,j_K} Q_{jr}\delta \ln x_j\;,
 \qquad
r =j_1,j_2,..,j_K\;,
\end{eqnarray}
%
where the coefficients $P_{ir}$ and $Q_{jr}$  involve the different
logarithmic derivatives $\alpha_{ij}$.
It follows from the preceding equation
that the  1$\sigma$ relative uncertainty in $x_r$ is given by
\begin{eqnarray}
\Delta \ln x_r &=&
\sqrt{
\sum_{i=i_1,..,i_K} P_{ir}^2 \left(\Delta \ln \phi_i
\right)^2
+ \sum_{j\neq j_1,..,j_K} Q_{jr}^2\left(\Delta \ln x_j^0
\right)^2} \nonumber \\
{\rm or,}~~~~
\ln \left(1+\frac{\Delta x_r}{x_r}\right) &=&
\sqrt{
\sum_{i=i_1,..,i_K} P_{ir}^2 
\left[\ln\left(1+\frac{\Delta \phi_i}{\phi_i}\right)\right]^2
+ \sum_{j\neq j_1,..,j_K} Q_{jr}^2
\left[\ln\left(1+\frac{\Delta x_j^0}{x_j^0}\right)\right]^2} 
\nonumber \\
{\rm or}~~~~
\frac{\Delta x_r}{x_r} 
&=& \left|
1 - \exp\left(
\sqrt{
\sum_{i=i_1,..,i_K} P_{ir}^2 
\left[\ln\left(1+\frac{\Delta \phi_i}{\phi_i}\right)\right]^2
+ \sum_{j\neq j_1,..,j_K} Q_{jr}^2
\left[\ln\left(1+\frac{\Delta  x_j^0}{x_j^0}\right)\right]^2} 
\right)\right| \qquad
\label{genuncert}
\end{eqnarray}
%
where $\Delta\phi_i/\phi_i$ is the 1$\sigma$ relative
uncertainty in the directly measured flux $\phi_i$ and 
$\Delta x_j^0/x_j^0$ is the 1$\sigma$ relative
uncertainties of the parameter $x_j^0$ as estimated in the SSM.  
\begin{table}
\begin{center}
\begin{tabular}{|c|c|c|c|c|c|c|c|c|}
\hline
&&&&&&&& \\
$j$ & 
$\alpha_{pp,j}$ & $\alpha_{pep,j}$ & $\alpha_{hep,j}$ & $\alpha_{Be,j}$
& $\alpha_{B,j}$ & $\alpha_{N,j}$ & $\alpha_{O,j}$ & 
$\alpha_{F,j}$ \\
&&&&&&&& \\
\hline
$S_{11}$ & $+0.14$ &$ -0.17$ & $-0.08$ & $-0.97$ & $-2.59$ & $-2.53$& $-2.93$& $-2.94$\\
$S_{33}$ & $+0.03$ & $+0.05$ & $-0.45$ & $-0.43$ & $-0.40$ & $+0.02$& $+0.02$& $+0.02$\\
$S_{34}$ & $-0.06$ & $-0.09$ & $-0.08$ & $+0.86$ & $+0.81$ & $-0.05$& $-0.05$& $-0.05$\\
$S_{1,14}$ & $-0.0$2 & $-0.02$ & $-0.01$ &  $   0 $& $+0.01$ & $+0.85$& $+1.00$& +0.01\\
$S_{17}$  & $0$ & $0$   & $0 $  &   $ 0$ & $+1.00$ &     0&     0&    0\\
$L_\odot$  & $+0.73$ & $+0.87$ & $+0.12$ & $+3.40$ & $+6.76$ & $+5.16$& $+5.94$&$ +6.25$\\
$Z/X$ & $-0.08$ &$-0.17$ &$-0.24$ & $+0.62$& $+1.36$ & $+1.99$ & $+2.06$&$+2.17$\\
$\tau_\odot$ & $-0.07$ & $0$  & $-0.11$  & $+0.69$ & $+1.28$ & $+1.01$& $+1.27$& $+1.29$\\
$O_\odot$  & $+0.14$ & $+0.24$ &$+0.54 $ &$ -1.49$ & $-2.93$ &$-1.81$ &$-2.25$ &$-2.35$ \\
$D_\odot$  & $+0.13$ & $+0.22$ &$+0.38 $ & $-0.96$ & $-2.20$ &$-2.86$& $-3.10$ &$-3.22$\\
$S_{e^-7}$   & $0$     & $0$    &$0$ & $ 0 $   &$ -1.00$ & $0$& $0$& $0$\\
\hline
\end{tabular}
\end{center}
\caption{\label{tab_alpha} Values of the logarithmic derivatives 
$\alpha_{ij}=\frac{\partial \ln \phi_i}{\partial \ln x_j}$,
corresponding to different 
solar neutrino fluxes $\phi_i$ ($pp$, $pep$, $hep$, $^7Be,j$,
$^8B$, $^{13}N$, $^{15}O$, $^{17}F$) and
the input parameters of the solar model (from ref. \cite{bp04}).
}
\end{table}

%
\section{Determining One SSM Input 
Parameter Using Measured $^8B$ Neutrino Flux}
%
\label{oneparam}

Following the technique described in Section\ \ref{nparam}, we can use
the measured value of the $^8B$ neutrino flux,
$\phi_{B}$, and its uncertainty, as given
in eq.\ (\ref{phiglobal}), to determine one of the parameters $x_{j_1}$
of the solar model together with its uncertainty,
$\frac{\Delta x_{j_1}}{x_{j_1}}$. They are given 
respectively by eqs. (\ref{xgensol}) and (\ref{genuncert}) 
reduced to K = 1:
\begin{eqnarray}
x_{j_1}
&=& x_{j_1}^0~
\left(\frac{\phi_{B}}{\phi_{B}^{SSM}}\right)^
{\frac{1}{\alpha_{B,j_1}}} \;, 
\label{xj1central}
\\
\frac{\Delta x_{j_1}}{x_{j_1}}
&=&
\left|
1 - \exp\left(
\sqrt{\left(\frac{1}{\alpha_{B,j_1}}\right)^2
\left[\ln\left( 1+ \frac{\Delta\phi_{B}}{\phi_{B}}\right)\right]^2
+
\sum_{j\neq j_1}
\left(\frac{\alpha_{B,j}}{\alpha_{B,j_1}}\right)^2
\left[\ln\left(1+\frac{\Delta x_j^0}{x_j^0}\right)\right]^2}\right)
\right|
%
\label{xj1error}
\end{eqnarray}
%
We show results for the central value of the input 
parameters with respect to their values used in BP04 SSM 
and their corresponding uncertainties in Table \ref{tab_1param}. 
The extreme right column of Table\ \ref{tab_1param} 
gives the factor by which the central values of
different parameters will change with respect to their 
values as used in the BP04 SSM \cite{bp04} 
if we use the measured $^8B$ neutrino 
flux, given in eq. (\ref{phiglobal}), for their calculations.  
Note that if the measured mean value of 
$\phi_{B}$ differs from the value predicted by the SSM,
$\phi_{B}^{SSM}$, the value of the parameter $x_{j_1}$ 
obtained using eq. (\ref{xj1central}) 
would differ from its SSM predicted value $x_{j_1}^0$ 
by a factor controlled by the corresponding logarithmic
derivative 
$\alpha_{B,j_1}=\frac{\partial \ln \phi_{B}}{\partial  \ln x_{j_1}}$.  

We present in Table  \ref{tab_1param} 
the uncertainty of each of 
the SSM input parameters, evaluated 
from eq. (\ref{xj1error}), for three
different ``benchmark'' values of the uncertainty 
in the measured $^8B$ neutrino flux. 
For comparison we have also included in column 5 of  
Table  \ref{tab_1param} the estimated uncertainties 
in the SSM parameters in the BP04 SSM.
It follows from  eq. (\ref{xj1error}) that the relative
uncertainty in the parameter $x_{j_1}$  
depends on the inverse power of the
magnitude of the corresponding logarithmic derivative 
$\alpha_{B,j_1}$. It should be clear from eq. (\ref{PhiB})
and Table \ref{tab_alpha} 
that since the relevant logarithmic derivatives
in the cases of the $S$-factors $S_{33}$ 
and  $S_{1,14}$ 
are relatively small, these quantities
cannot be determined with smaller uncertainties
by using even a high precision measurement of the $^8$B 
neutrino flux. We therefore do not show the results for
these cases in Table \ref{tab_1param}.  
It follows from the results reported in 
Table \ref{tab_1param}, in particular, that the SSM
parameters like  $S_{11}$, $Z/X$, $L_\odot$ and $O_{\odot}$ 
can be determined with uncertainties less 
than 10\% owing to the relatively large values
of their corresponding logarithmic derivatives. 
For the uncertainty in diffusion parameter $D_\odot$ 
we get approximately 10.6\%.
The table shows also that using a direct high precision 
measurement of the $^8B$ neutrino flux
can allow to determine the
parameter $Z/X$ with an uncertainty which is smaller
than its currently estimated uncertainty in the 
BP04 SSM \cite{bp04}. At the same time,
the parameter $S_{17}$ is determined much less
precisely - with uncertainty of 25\%, 
in spite of the fact that $\phi_B \propto S_{17}$.

\begin{figure}
\epsfig{file=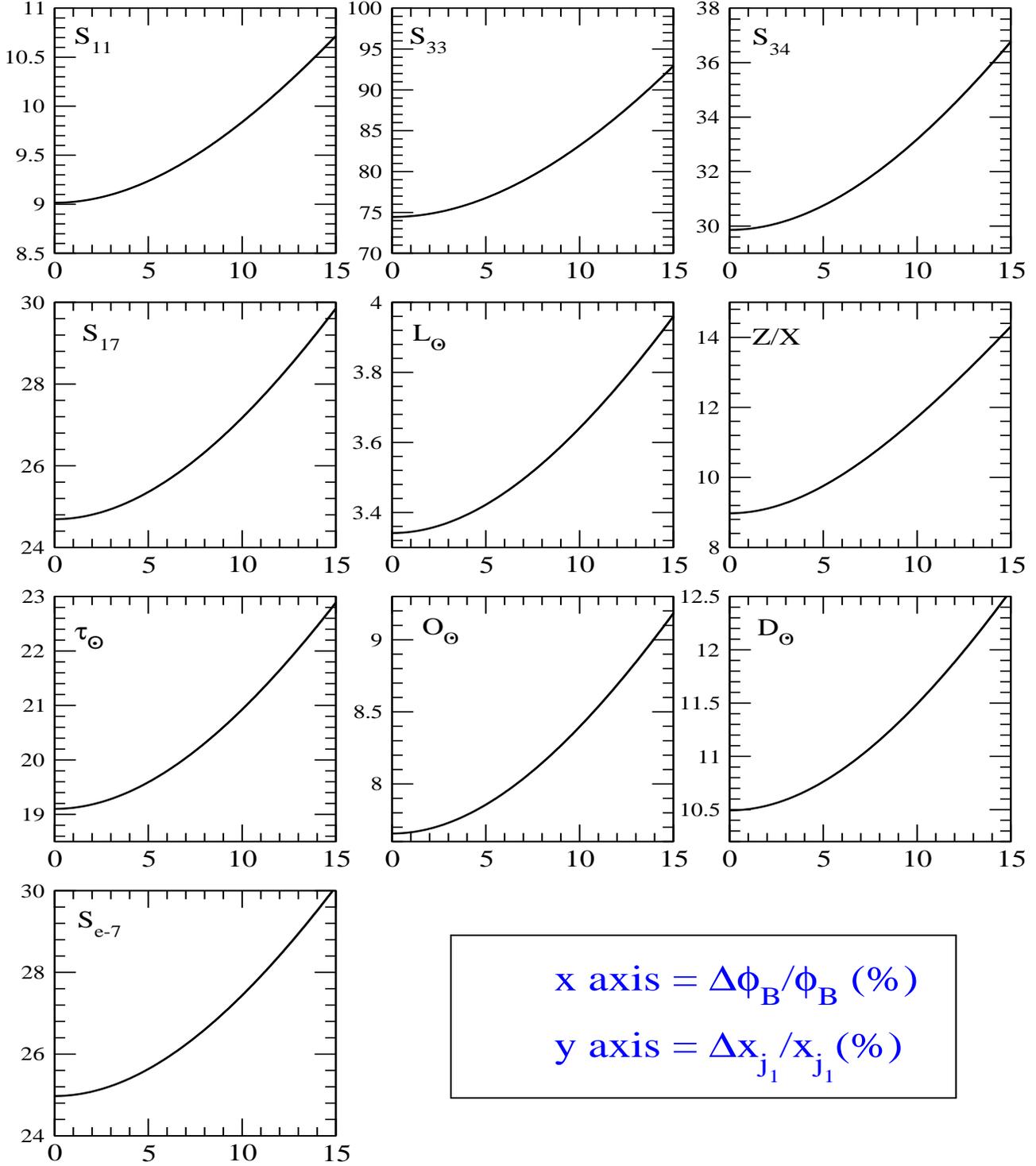,width=7in,height=8in}
\caption{The 1$\sigma$
fractional uncertainty ($\frac{\Delta x_{j_1}}{x_{j_1}}$)
of the various SSM input parameters 
as a function of the uncertainty in the measured $^8B$ neutrino flux 
($\Delta\phi_{B}/\phi_{B}$).
\label{fig1}
}  
\end{figure}

We see from Table \ref{tab_1param} that the uncertainties 
of most of the SSM parameters under discussion
are essentially stable when the 1$\sigma$ error in 
the measured $^8B$ neutrino flux
changes from 2\% to 4\%. 
The reason for such a behavior is that for very small 
values of $\Delta \phi_B/\phi_B$, the first term in 
the right hand side of eq. (\ref{xj1error}) is 
much smaller than the second term which 
controls the uncertainty  $\Delta x_{j_1}/x_{j_1}$. 
In Fig. \ref{fig1} we have plotted the fractional 
uncertainty in each of the SSM parameters 
as a function of the 1$\sigma$ error in 
the measured value of the $^8{B}$ neutrino flux
\footnote{Obviously, the uncertainty in the value of 
a given SSM input parameter determined 
exploiting a solar neutrino flux measurement  
depends also on the uncertainties in the remaining 
SSM input parameters used in the evaluation.}.   
Figure \ref{fig1} shows that if the $^8B$ flux 
uncertainty is larger than $\sim$ 5\%, 
the first term in eq. (\ref{xj1error})
would dominate over the second
and $\Delta x_j/x_j$ can exhibit a stronger 
dependence on the uncertainty $\Delta\phi_{B}/\phi_{B}$.
The degree of this dependence is controlled by
the corresponding $\alpha_{Bj_1}$ value. However, 
as we have shown before 
(cf. Fig. \ref{fig:fb} and Table \ref{tab:fb}),
the uncertainty in the value of $f_B$, 
determined from the current data, is already 
approximately  4\%. The second term in 
eq. (\ref{xj1error}) is dominant and 
we do not expect any significant improvement 
of the precision of determination  of the 
SSM input parameters with 
more precise measurement of $\phi_B$ alone. 
Small improvements can nonetheless be expected, 
especially in what concerns the precision of 
determination of $Z/X$.

%
\begin{table}
\begin{center}
\begin{tabular}{|c|c|c|c|c|c|}
\hline
\multirow{2}{30mm}{Name of model parameters ($x_{j_1}$)} & 
\multicolumn{3}{|c|}{$\frac{\Delta x_{j_1}}{x_{j_1}}$ (\%)} &  
\multirow{2}{20mm}{$\hfill \frac{\Delta x_{j_1}^0}{x_{j_1}^0}$
  (\%)\hfill} & 
\multirow{2}{15mm}{\hfill $\frac{x_{j_1}}{x_{j_1}^0}$ \hfill}\\
\cline{2-4}
& $\frac{\Delta\phi_{B}}{\phi_{B}}$ = 2\%
& $\frac{\Delta\phi_{B}}{\phi_{B}}$ = 3\%
& $\frac{\Delta\phi_{B}}{\phi_{B}}$ = 4\% 
& & \\
\hline
$S_{11}$            &   9.05 &   9.10  &  9.16   & 0.4  & 1.07\\
$S_{34}$            &  30.01 &  30.19  &  30.44  & 9.4  & 0.81\\
$S_{17}$            &  24.80 &  24.94  &  25.12  & 3.8  & 0.84\\
Luminosity $L_\odot$        &   3.35 &   3.37  &   3.39  & 0.4  & 0.97\\
Z/X                 &   9.11 &   9.27  &   9.48  & 15.0 & 0.88\\
Age of sun          &  19.18 &  19.28  &  19.42  & 0.4  & 0.87\\
Opacity $O_\odot$   &   7.69 &   7.73  &   7.79  & 2.0  & 1.06\\
Diffusion $D_\odot$ &  10.54 &  10.59  &  10.66  & 2.0  & 1.08\\
$S_{e^-7}$            &  25.08 &  25.22  &  25.40  & 2.0  & 1.19\\
\hline
\end{tabular}
\caption{\label{tab_1param} The uncertainties for each of the
  SSM input parameters, which are expected to be obtained 
  if a high precision direct measurement of the $^8B$ neutrino flux is 
  used to determine the corresponding SSM parameter. Given are also
  the uncertainties in the SSM input parameters \cite{bp04}, 
  used in the SSM calculations  of the solar neutrino fluxes.}
\end{center}
\end{table}

%
\section{Determining Two SSM Input Parameters
Using Measured $^8B$ and $^7Be$ Neutrino Fluxes}
\label{twoparam}
%
%

  As we have discussed in Section 3, 
a relatively high precision 
measurement of the \ber neutrino flux can be 
performed by the Borexino experiment.
The \kl experiments can also provide valuable data
on  $\phi_{Be}$.  One could use the 
experimentally measured values of the
$^8B$ and $^7Be$ neutrino fluxes to determine any 
two of the SSM input parameters  
\footnote{We have also examined the uncertainties of
the SSM parameters one obtains if only high precision 
prospective data on the $^7Be$ neutrino flux is used
to determine the parameters. We have found, in particular, 
that $S_{33}$ and $S_{34}$ can be determined 
with a precision of 
43.3\% (33\%) and 17.2\% (12\%), respectively, if 
the \ber neutrino flux is measured
with 1$\sigma$ error of 10\% (2\%).
This should be compared with the uncertainties of
76\% and 30\% in $S_{33}$ and $S_{34}$
we have obtained in the previous 
Section, using \br neutrino 
flux measurement with 4\% uncertainty.
Even if we take the 1$\sigma$ error in
$\phi_{Be}$ to be 2\%,
all the other SSM parameters are detrmined 
with uncertainties which are larger
than those we found in Section 5
when the same parameters
are determined from the measured $\phi_{B}$
with 1$\sigma$ error of 4\%.}.   
Equations (\ref{xgensol}) and (\ref{genuncert}) for K=2
give the central values and the uncertainties for any two parameters 
$x_{j_1}$ and $x_{j_2}$ as 
\begin{eqnarray}
x_{j_1}
&=&
x_{j_1}^0\times
\left[
\left(\frac{\phi_{B}}{\phi_{B}^{SSM}}\right)^{\alpha_{Be,j_2}}
\times 
\left(\frac{\phi_{Be}}{\phi_{Be}^{SSM}}\right)^{-\alpha_{B,j_2}}
\right]
^{\left(
{\frac{1}{\alpha_{B,j_1}\alpha_{Be,j_2} 
- \alpha_{Be,j_1}\alpha_{B,j_2}}}\right)}\;,
\label{xj1central_2}
\\
x_{j_2}
&=&
x_{j_2}^0\times
\left[
\left(\frac{\phi_{B}}{\phi_{B}^{SSM}}\right)^{\alpha_{Be,j_1}}
\times 
\left(\frac{\phi_{Be}}{\phi_{Be}^{SSM}}\right)^{-\alpha_{B,j_1}}
\right]
^{\left(
{\frac{1}{\alpha_{B,j_2}\alpha_{Be,j_1} 
- \alpha_{Be,j_2}\alpha_{B,j_1}}}\right)}\;,
\label{xj2central_2}
\end{eqnarray}
and 
\begin{eqnarray}
\left(\frac{\Delta x_{j_1}}{x_{j_1}}\right) \!\!=\!\!
\left|1\!-\!\exp\left[
\frac{
\left(\frac{1}{\alpha_{B,j_2}}\right)^2 
\Delta (\ln\phi_B)^2 +
\left(\frac{1}{\alpha_{Be,j_2}}\right)^2
\Delta (\ln\phi_{Be})^2
+{\sum\atop{j\neq j_1,j_2}}
(\frac{\alpha_{B,j}}{\alpha_{B,j_2}}
 - \frac{\alpha_{Be,j}}{\alpha_{Be,j_2}})^2
\Delta (\ln x_j^0)^2}
{\left(\frac{\alpha_{B,j_1}}{\alpha_{B,j_2}}
-\frac{\alpha_{Be,j_1}}{\alpha_{Be,j_2}}\right)^2}
\right]^{\frac{1}{2}}\right| 
\label{xj1error_2} 
\end{eqnarray}
\begin{eqnarray}
\left(\frac{\Delta x_{j_2}}{x_{j_2}}\right) \!\!=\!\!
\left|1\!-\!\exp\left[
\frac{
\left(\frac{1}{\alpha_{B,j_1}}\right)^2 
\Delta (\ln\phi_B)^2
+\left(\frac{1}{\alpha_{Be,j_1}}\right)^2
\Delta (\ln\phi_{Be})^2
+{\sum\atop{j\neq j_1,j_2}}
(\frac{\alpha_{B,j}}{\alpha_{B,j_1}}
 - \frac{\alpha_{Be,j}}{\alpha_{Be,j_1}})^2
\Delta (\ln x_j^0)^2}
{\left(\frac{\alpha_{B,j_2}}{\alpha_{B,j_1}}
-\frac{\alpha_{Be,j_2}}{\alpha_{Be,j_1}}\right)^2}
\right]^{\frac{1}{2}}\right| 
\label{xj2error_2} 
\end{eqnarray}

%

 Equations (\ref{xj1central_2}) and 
(\ref{xj2central_2}) imply, in particular, that
the values of the parameters $x_{j_1}$ and $x_{j_2}$
could differ from their respective SSM values  
by factors determined by 
the four logarithmic derivatives - 
$\alpha_{B,j_1}$, $\alpha_{B,j_2}$, $\alpha_{Be,j_1}$ and
$\alpha_{Be,j_2}$.
For calculating the SSM parameter uncertainties, 
we take all possible combinations
of $\{x_{j_1}$,$x_{j_2}\}$ and use 
eqs. (\ref{xj1error_2}) and (\ref{xj2error_2}) to get
the corresponding errors on these sets of two parameters, 
assuming a measurement of $^8B$ and $^7Be$
neutrino fluxes respectively with 4\% and 6\% uncertainty 
\footnote{While $\phi_B$ is already known with a 
4\% uncertainty, $\phi_{Be}$ could be determined,
as we have discussed above,
with a 6\%  error by the Borexino experiment
(see ref. \cite{roadmap} for details).}.
The uncertainties of $x_{j_1}$ and $x_{j_2}$, as given in
eqs. (\ref{xj1error_2}) and (\ref{xj2error_2}), are controlled
in a rather complicated way
by both the magnitude and the 
relative signs of the different logarithmic
derivatives. However, it follows from eqs. 
(\ref{xj1error_2}) and (\ref{xj2error_2})
that  if for a certain pair of SSM parameters $x_{j1}$ and $x_{j2}$
the relation $\alpha_{B,j_2}/\alpha_{B,j_1}\cong
\alpha_{Be,j_2}/\alpha_{Be,j_1}$ holds, 
these parameters would be determined with 
poor accuracy even if one uses high 
precision data on the $^8B$ and $^7Be$ 
neutrino fluxes. Table \ref{tab_alpha} suggests that 
such pairs can be, for instance,  $\{S_{33},S_{34}\}$,
$\{L_\odot,O_\odot\}$ and $\{Z/X,D_\odot\}$. 

  Among the solar physics parameters 
the opacity $O_\odot$ can be determined with a
9\% uncertainty in pair with $S_{34}$, or 
$S_{17}$, or $S_{e^-7}$, for which we get
at the same time uncertainty of approximately 16\%.
For the diffusion $D_\odot$ we find an uncertainty of 11\%
when determined in combination with
$S_{33}$ or $S_{34}$. In these cases, however,
$S_{33}$ and $S_{34}$ are found within
36\% and 14\%, respectively. 

   In Table \ref{tab_combination} we present results 
only for those combinations of two SSM parameters which
are determined with an uncertainty smaller than 15\%
using directly measured values of $\phi_B$ and $\phi_{Be}$. 
It is interesting to note 
from Table \ref{tab_combination} that 
the set $\{S_{34}$,$Z/X\}$ can be determined 
with a relatively good precision:
the predicted uncertainty of $Z/X$ is smaller than 
the estimated one within the BP04 SSM \cite{bp04},
while the uncertainty in $S_{34}$ is approximately 14.3\%. 
The latter should be compared with the 1$\sigma$ error
of 9.4\%, quoted when $S_{34}$ is obtained using 
the data on the reaction $^3He + ^4He \rightarrow ^7Be + \gamma$.
The parameter $S_{34}$ can be determined with 12.7\% 
uncertainty in the combination $\{S_{34},S_{11}\}$.
As can be seen by comparing Table \ref{tab_1param} with 
Table \ref{tab_combination}, one gets a somewhat better 
precision on $Z/X$ if the value of $Z/X$ is obtained from the
experimental information on the $^8B$ neutrino flux only
\footnote{The reason for this 
behavior can be traced to the complicated nature of the 
eqs. (\ref{xj1error_2}) and (\ref{xj2error_2}).}. 
However, the uncertainty in the determination of
$S_{34}$ can be much smaller when 
in addition to the data on the $^7Be$ neutrino flux 
one uses the data on the $^8B$ neutrino flux as well.

\begin{table}[t]
\begin{center}
\begin{tabular}{|c|c||c|c|}
\hline
\multicolumn{2}{|c||}{combinations} & \multirow{2}{25mm}{\hfill
  $\frac{\Delta x_{j_1}}{x_{j_2}}$ (\%)\hfill} & 
\multirow{2}{25mm}{\hfill$\frac{\Delta x_{j_2}}{x_{j_2}}$ (\%)\hfill} \\
\cline{1-2} $j_1$  & $j_2$ &&\\
\hline
$S_{34}$    &    $S_{11}$     & 12.71    &  8.81\\  
$S_{34}$    &    $Z/X$        & 14.25    & 12.26\\  
$S_{34}$    &    $D_\odot$    & 13.79    & 11.35\\  
\hline
\end{tabular}
\end{center}
\caption{\label{tab_combination} 
 The list of different combinations of two SSM parameters
 which are determined with an uncertainty smaller than 
 15\% using data of prospective direct measurements of 
  $^8B$ and $^7Be$ 
neutrino fluxes with 1$\sigma$ errors of 4\% and 6\%, respectively.  
The relative uncertainties on the SSM parameters 
thus determined are also given.
}
\end{table}
%
\begin{table}[t]
\begin{center}
\begin{tabular}{|c|c||c||c|}
\hline
$\frac{\Delta \phi_B}{\phi_B}$ (\%)& $\frac{\Delta
  \phi_{Be}}{\phi_{Be}}$(\%)  
& $\frac{\Delta (Z/X)}{(Z/X)}$ (\%)& $\frac{\Delta S{34}}{S_{34}}$ (\%)\\
\hline
&&&\\
2   & 2  & 8.81  & 6.85  \\
    & 4  & 9.83  & 10.02 \\
    & 6  & 11.32 & 13.83 \\
    & 8  & 13.11 & 17.93 \\
    & 10 & 15.10 & 22.21\\
&&&\\
3   & 2  & 9.30  & 7.16  \\
    & 4  & 10.29  & 10.24 \\
    & 6  & 11.72 & 14.00 \\
    & 8  & 13.47 & 18.07 \\
    & 10 & 15.42 & 22.34 \\
&&&\\
4   & 2  & 9.96 &  7.60 \\
    & 4  & 10.88 & 10.56 \\
    & 6  & 12.26 & 14.25 \\
    & 8  & 13.95 & 18.27 \\
    & 10 & 15.85 & 22.51 \\
&&&\\
\hline 
\end{tabular}
\end{center}
\caption{\label{tab_2param} 
Relative uncertainties of the 
set of two parameters - $Z/X$ and $S_{34}$, 
determined using  prospective high precision 
data on the 
$^8B$ and $^7Be$ 
neutrino fluxes. The results correspond 
to different sets of assumed 1$\sigma$ errors 
in the measured values of the two fluxes.}
\end{table}
%
\begin{figure}
\epsfig{file=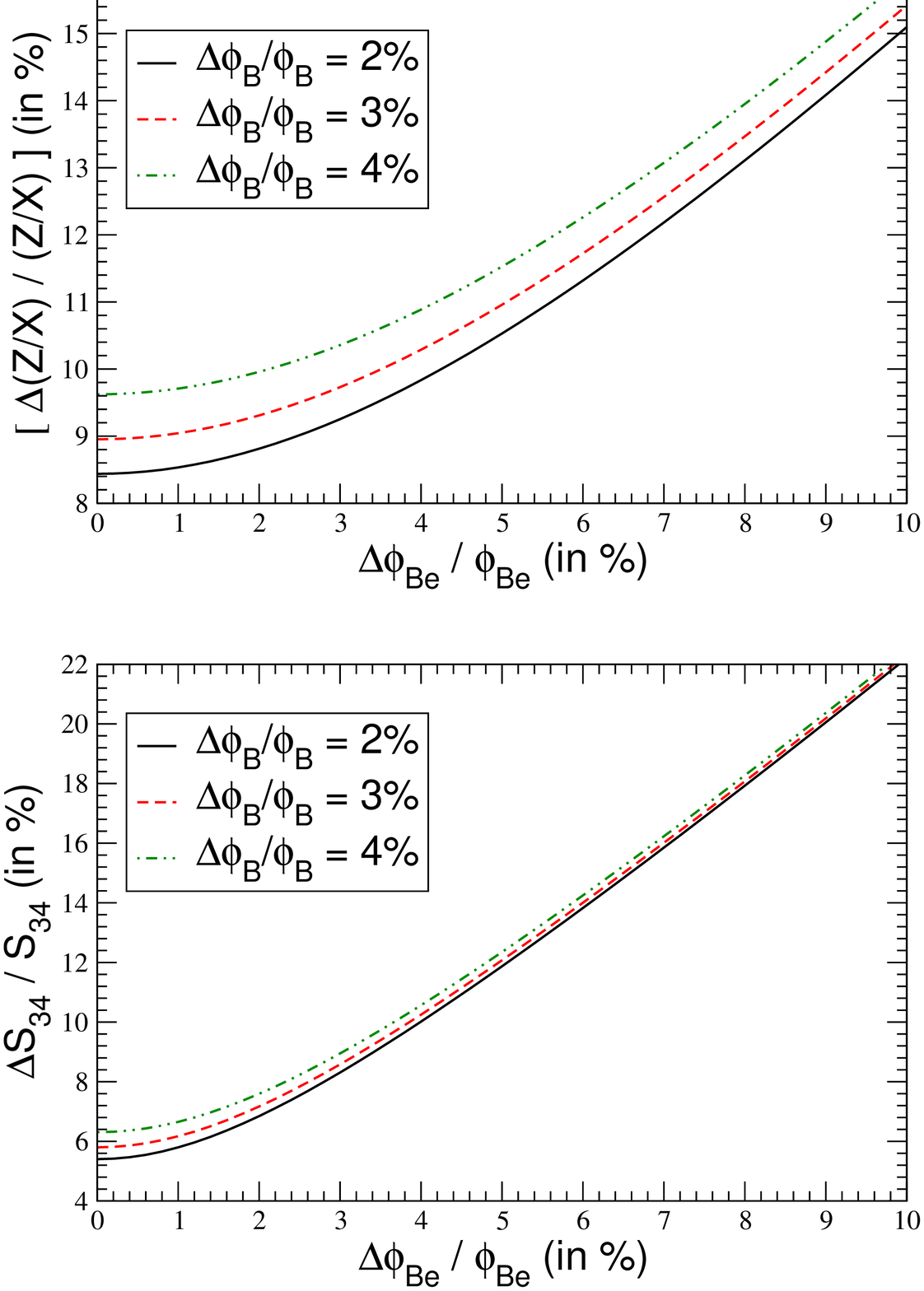,width=7in,height=8in}
\caption{The 1$\sigma$ fractional uncertainties of 
the SSM input parameters $S_{34}$ and $Z/X$, 
determined from data on the $^8B$ and $^7Be$ 
neutrino fluxes, as a function of the uncertainty in the measured $^7Be$ 
neutrino flux, for three different values of the 
1$\sigma$ error in the measured $^8B$ neutrino flux. 
\label{fig:combination}
}  
\end{figure}

  In Table \ref{tab_2param} we
present results on the uncertainties of the 
parameter combination $\{Z/X$,$S_{34}\}$, determined 
from data on the $^8B$ and $^7Be$ neutrino fluxes.
The results correspond to
1$\sigma$ errors of 2\%, 3\% and 4\%
in the measured value of the
\br neutrino flux,
and of 2\%, 4\%, 6\%, 8\% and 10\%
in the measured value of $\phi_{Be}$. 
We observe that if the 1$\sigma$ uncertainties in
the values of $\phi_B$ and $\phi_{Be}$ 
are smaller than 4\%,
the uncertainty in the value of $S_{34}$,  
determined using the neutrino flux measurements, 
would be smaller than its present
uncertainty of $\sim$ 9\% . 
The uncertainty in the determination of 
$Z/X$ would be smaller than its currently
SSM estimated one of 15\% 
if $\phi_{Be}$ is measured with an 
error not exceeding approximately 
$10\%$ at 1$\sigma$.

  In Fig. \ref{fig:combination} we show the 
expected 1$\sigma$ uncertainties of the pair 
of parameters $\{S_{34},Z/X\}$,
$\Delta S_{34}/S_{34}$ and $\Delta (Z/X)/(Z/X)$,
as continuous functions of the uncertainty 
in the measured $^7Be$ neutrino 
flux, $\Delta \phi_{Be}/\phi_{Be}$,
for three different values for the uncertainty 
in the measured $^8B$ neutrino flux,
$\Delta \phi_{B}/\phi_{B}$.
The figure shows that the dependence of 
$\Delta S_{34}/S_{34}$ on 
$\Delta \phi_{Be}/\phi_{Be}$
is stronger than  that on 
$\Delta \phi_{B}/\phi_{B}$,
while $\Delta (Z/X)/(Z/X)$ depends on the accuracy of 
measurement of both fluxes. This feature is due to the fact that
$\alpha_{Be,Z/X} \simeq 2.2\times \alpha_{B,Z/X}$, while
$\alpha_{Be,S_{34}} \simeq \alpha_{B,S_{34}}$.

%
\section{Determining Three SSM Input Parameters Using
Data on the $^8B$, $^7Be$ and  $pp$ Neutrino Fluxes}
\label{threeparam}

As we have already discussed in Section 3,
the $pp$ neutrino flux, $\phi_{pp}$, is 
determined with an uncertainty of 2\%
from the present solar and reactor neutrino data 
if one employs the luminosity constraint. 
Future solar neutrino experiments can 
provide a remarkably precise measurement of 
$\phi_{pp}$ \cite{roadmap}.
In this Section we use high precision prospective 
data on \br, \ber and $pp$ neutrino 
fluxes to simultaneously determine 
any three of the SSM input parameters.

 Equations\ (\ref{xgensol}) and\ (\ref{genuncert}) for K=3 give 
the central values and uncertainties for any three parameters 
$x_{j_1}$, $x_{j_2}$ and $x_{j_3}$ as
\begin{eqnarray}
x_r 
&=& 
\left[
\left(\frac{\phi_{pp}}{\phi_{pp}^{SSM}}\right)^{\beta_{pp,r}}
\left(\frac{\phi_{Be}}{\phi_{Be}^{SSM}}\right)^{\beta_{Be,r}}
\left(\frac{\phi_{B}}{\phi_{B}^{SSM}}\right)^{\beta_{B,r}}
\right]^{1/Det A}
\times 
x_r^0, \qquad (r=j_1,j_2,j_3)
\label{xvalue3}
\end{eqnarray}
\begin{center}
and
\end{center}
{\small
\begin{eqnarray}
&&\left[\ln \left(1+\frac{\Delta x_r}{x_r}\right)\right]^2 =
\nonumber \\
&&
\left(\frac{1}{{\rm Det A}}\right)^2 \times
\bigg[
\beta_{pp,r}^2 
\left\{\ln\left(1+\frac{\Delta\phi_{pp}}{\phi_{pp}}\right)\right\}^2
+
\beta_{Be,r}^2 
\left\{\ln\left(1+\frac{\Delta\phi_{Be}}{\phi_{Be}}\right)\right\}^2
+
\beta_{B,r}^2 
\left\{\ln\left(1+\frac{\Delta\phi_{B}}{\phi_{B}}\right)\right\}^2 
\nonumber \\
&&
+\sum_{j\neq j_1,j_2,j_3}
(\beta_{pp,r}\alpha_{pp,j}+\beta_{Be,r}\alpha_{Be,j}+
\beta_{B,r}\alpha_{B,j})^2
\left\{\ln\left(1+\frac{\Delta x_j}{x_j}\right)\right\}^2 
\bigg]^{},  \quad (r = j_1,j_2,j_3)
\label{xerror3}
\end{eqnarray}
}
%
where A is 3$\times$3 matrix given by
\begin{eqnarray}
A &=& 
\pmatrix{ 
\alpha_{pp,j_1} & \alpha_{pp,j_2} & \alpha_{pp,j_3} \cr 
\alpha_{Be,j_1} & \alpha_{Be,j_2} & \alpha_{Be,j_3} \cr
\alpha_{B,j_1} & \alpha_{B,j_2} & \alpha_{B,j_3}
} .
\end{eqnarray}
%
$\beta_{pp,r}$, $\beta_{Be,r}$ and $\beta_{B,r}$ are the cofactors
of the matrix elements $\alpha_{pp,r}$, $\alpha_{Be,r}$ and 
$\alpha_{B,r}$ respectively ($r=j_1,j_2,j_3$).
%
\begin{table}[t]
\begin{center}
\begin{tabular}{|c|c|c||c|c|c|}
\hline
\multicolumn{3}{|c||}{combinations} & \multirow{2}{25mm}{\hfill
  $\frac{\Delta x_{j_1}}{x_{j_2}}$ (\%)\hfill} & 
\multirow{2}{25mm}{\hfill$\frac{\Delta x_{j_2}}{x_{j_2}}$ (\%)\hfill} & 
\multirow{2}{25mm}{\hfill$\frac{\Delta x_{j_3}}{x_{j_3}}$(\%)\hfill}\\
\cline{1-3} $x_{j_1}$  & $x_{j_2}$ & $x_{j_3}$ & & &\\
\hline
$S_{11}$ & $S_{34}$  & $L_\odot$    & 6.16  & 7.98 & 1.62\\ 
$S_{11}$ & $L_\odot$ & $O_\odot$    & 13.67 & 1.72 & 12.25\\
$S_{34}$ & $L_\odot$ & $Z/X$        & 8.87  & 1.00 & 8.63\\ 
$S_{34}$ & $L_\odot$ & $O_\odot$    & 10.45 & 1.69 & 5.31\\ 
$S_{34}$ & $L_\odot$ & $D_\odot$    & 9.03  & 1.57 & 5.77\\ 
$S_{34}$ & $L_\odot$ & $\tau_\odot$ & 11.40 & 1.66 & 12.86\\
$S_{11}$ & $S_{17}$  & $L_\odot$    & 4.73  & 11.05 & 1.73\\ 
$S_{11}$ & $L_\odot$ & $S_{e^-7}$   & 4.53  & 1.73  & 11.05\\
$S_{33}$ & $S_{17}$  & $L_\odot$    & 12.40 & 8.47  & 1.95\\ 
$S_{33}$ & $L_\odot$ & $S_{e^-7}$   & 13.43 & 1.95  & 8.47\\
$S_{34}$ & $S_{17}$  & $L_\odot$    & 8.61  & 8.98  & 1.86\\
$S_{34}$ & $L_\odot$ & $S_{e^-7}$   & 6.33  & 1.85  & 8.98\\ 
$S_{17}$ & $L_\odot$ & $Z/X$        & 9.26  & 1.21  & 6.65\\
$S_{17}$ & $L_\odot$ & $\tau_\odot$ & 9.05  & 1.85  & 7.34\\
$S_{17}$ & $L_\odot$ & $O_\odot$    & 9.21  & 1.84  & 3.40\\
$S_{17}$ & $L_\odot$ & $D_\odot$    & 9.74  & 1.74  & 4.68\\
$L_\odot$&  $Z/X$   & $S_{e^-7}$    & 1.21  & 6.65  & 9.88\\
$L_\odot$&$\tau_\odot$& $S_{e^-7}$  & 1.85  & 7.42  & 9.29\\
$L_\odot$&$O_\odot$& $S_{e^-7}$  & 1.84  & 3.40  &9.48\\
$L_\odot$&$D_\odot$& $S_{e^-7}$  & 1.73  & 4.68  &10.12\\
\hline
\end{tabular}
\end{center}
\caption{\label{tab_combination3} The fractional uncertainties in
different combinations of the SSM input parameters 
for which we get uncertainty smaller than 15\%,  
assuming  3\%, 4\% and 1\% errors in the 
``measured''  $\phi_B$, $\phi_{Be}$ and $\phi_{pp}$
neutrino fluxes, respectively. }
\end{table}
%
We take different combinations of three SSM 
parameters and calculate their uncertainties using 
3\%, 4\% and 1\% as illustrative 1$\sigma$ errors in the 
measured values of $^8B$, $^7Be$ and $pp$ neutrino 
fluxes, respectively. We find that for almost all 
solar model  parameters, 
the uncertainties reduce 
when using the combined information on the
$^7B$, $^8Be$ and $pp$ neutrino fluxes, compared to 
what we have obtained by  using 
only prospective data on $^8B$ and/or $^7Be$ fluxes. 
In Table\ \ref{tab_combination3} we present results 
only for those sets of three SSM input parameters 
which are determined with uncertainties smaller than
15\% each.  

As Table\ \ref{tab_combination3}
shows, the most precise determination of $S_{34}$ 
occurs in the combination $\{S_{34},L_\odot,S_{e^-7}\}$, 
while $Z/X$ is best determined in the sets 
$\{S_{17},L_\odot,Z/X\}$
and $\{S_{e^-7},L_\odot,Z/X\}$. For both these parameters 
the uncertainties we get are smaller 
than those in the respective BP04 SSM predictions. 

Note also that 
we get a rather accurate determination 
of $S_{17}$ from the combination 
$\{S_{34},S_{17},L_\odot\}$, and of $L_\odot$ from 
$\{S_{34},L_\odot,Z/X\}$. Although the uncertainty 
on $S_{17}$ thus obtained of $\sim$(8\% - 10\%) 
is larger than the currently estimated 
uncertainty in the value of
$S_{17}$ found from relevant 
nuclear reaction data (see Section 2), 
our results on $S_{17}$ can be used,
in particular, as a consistency check, 
e.g., of the extrapolation procedure
employed to get $S_{17}$ from the data.
In what concerns the other parameters, 
we get the best determination of $S_{11}$ 
from $\{S_{11},,L_\odot,S_{e^-7}\}$, of $D_\odot$ from 
$\{S_{17},L_\odot,D_\odot\}$, and of $\tau_\odot$ from 
$\{S_{17},L_\odot,\tau_\odot\}$. 

  We stress that even though the precision we get for 
$L_\odot$ is worse than the precision achieved in the
direct measurement of $L_\odot$, 
the method we used to determine $L_\odot$
allows to perform a fundamental test of the thermo-nuclear fusion 
theory of energy generation in the Sun, 
as well as to test the hypothesis that the Sun
is in an approximate steady state 
in what regards the energy produced 
in its central region and the 
energy emitted from its surface.

%
\section{Conclusions}
\label{summary}
%
%

    In the present article we have studied the 
possibility of using the
precision data (current and prospective) on the 
i) $^8$B, ii) $^8$B and $^7$Be, and 
iii) $^8$B, $^7$Be and $pp$, solar neutrino fluxes 
in order to obtain ``direct'' information 
(i.e., to constrain or determine) on at least 
some of the eleven solar model parameters -  
opacity, diffusion, heavy element surface abundance,
nuclear reaction $S$-factors, etc., which enter into the 
calculations of the fluxes in the 
Standard Solar Model (SSM). Our work was inspired 
by the remarkable  progress made in the studies 
of solar neutrinos in the last several years, 
which led to an unexpectedly precise 
determination of the solar neutrino 
oscillation parameters and of the 
$^8$B neutrino flux, as well as by 
the prospects for high precision 
measurements of the $^7$Be and $pp$ 
neutrino fluxes. It was stimulated 
also by the realization that
the solar physics parameters like the
opacity ($O_\odot$), diffusion ($D_\odot$) 
and heavy element surface abundance ($Z/X$),
can never be measured in direct experiments.
The solar photon luminosity $L_{\odot}$
is measured directly with very high accuracy. 
However, the ``conventionally'' 
measured luminosity of the Sun
is determined by photons
produced in the central region of the Sun,
which took $\sim 4\times 10^4$ years
to reach the surface of the Sun
from which they are emitted 
(see, e.g., \cite{Bahcall:2004mz}).
The luminosity determined 
from solar neutrino 
flux measurements provides
``real time'' information on the rates of 
nuclear fusion reactions in 
the central region of the Sun, 
in which the solar energy is generated:
neutrinos are simultaneously
produced in these reactions
with the photons observed in the form of 
solar luminosity, but
it takes solar neutrinos approximately only
8 minutes to reach the Earth. 
Similar considerations apply, 
perhaps to somewhat less extent,
to the $S$-factors $S_{11}$, $S_{33}$, 
$S_{34}$, $S_{1,14}$ and $S_{17}$,
directly related to the 
rates of the nuclear fusion reactions,
on which the SSM predictions for the solar 
neutrino fluxes depend and in which
the solar energy is generated.
They can be and are measured 
in direct experiments on Earth. 
However, this is done at energies which 
are considerably higher than the energies 
at which the reactions take place in the central 
part of the Sun. As a consequence, one has to 
employ an extrapolation procedure (based on 
nuclear theory) in order to obtain the values 
of the rates at the energy of interest,
corresponding to the physical conditions 
in the central part of the Sun.

  We have derived the basic equations for determining 
the central values of the SSM input parameters and 
their uncertainties using results of direct 
measurements of solar neutrino fluxes 
(Section 4, eqs. (\ref{xgensol}) and\ (\ref{genuncert})).
If we have $r$ measured solar neutrino fluxes,
at most $r$ SSM input parameters can be 
determined using 
the data on the solar neutrino fluxes.
For the remaining 
SSM parameters we have to use the 
SSM values and estimated uncertainties. 
All our numerical results are based 
on the predictions of the SSM of Bahcall 
and Pinsonneault from 2004 \cite{bp04}.
These include the dependence of different solar 
neutrino fluxes (\br, \ber, $pp$) 
on the SSM input parameters, 
and, whenever necessary, the predicted values of 
the SSM input parameters and their uncertainties.

  We used first the precise value 
of the \br neutrino flux, $\phi_B$, obtained from 
global analysis of solar neutrino and \kl data, 
to determine each of the SSM parameters on which $\phi_B$ depends. 
If the measured mean value of 
$\phi_{B}$ differs from the value predicted by the SSM,
$\phi_{B}^{SSM}$, the value of the parameter $x_{j_1}$ 
obtained using the data on $\phi_B$
would differ from its SSM predicted value $x_{j_1}^0$ 
by a factor controlled by the logarithmic derivative 
$\alpha_{B,j_1}=\frac{\partial \ln \phi_{B}}{\partial  \ln x_{j_1}}$
(see eq. (\ref{xj1central})).  
The relative uncertainty in the parameter $x_{j_1}$  
thus found depends on the inverse power of the
magnitude of
$\alpha_{B,j_1}$ (eq. (\ref{xj1error})). 
Since, according to the BP04 SSM, 
the relevant logarithmic derivatives
in the cases of the nuclear reaction 
$S$-factors $S_{33}$ and  $S_{1,14}$ 
are relatively small (see Table \ref{tab_alpha}), 
these quantities cannot be determined 
with sufficiently good accuracy even 
by using a high precision 
measurement of the $^8$B neutrino flux.
The results of this part of our analysis 
are summarised in Table \ref{tab_1param}.
We have found, in particular, 
that the SSM parameters like  $S_{11}$, 
$Z/X$, $L_\odot$ and $O_{\odot}$ 
can be determined with uncertainties less 
than 10\% owing to the relatively large values
of their corresponding logarithmic derivatives. 
For the uncertainty in diffusion parameter $D_\odot$ 
we get approximately 10.6\%.
Our results show that 
the parameter $Z/X$ can be determined 
with an uncertainty which is smaller
than its currently estimated uncertainty in the 
BP04 SSM \cite{bp04}. We have found also 
that the uncertainties of most of the SSM 
parameters under discussion
practically do not change when the 1$\sigma$ error in 
the measured $^8B$ neutrino flux
is reduced from 4\% to 2\%. 

 We have performed a similar analysis 
by combining a prospective high 
precision measurement of the \ber neutrino flux with 
the \br neutrino flux measurement. In this case 
it is possible to determine simultaneously two 
SSM input parameters, $x_{j_1}$ and $x_{j_2}$,
including their uncertainties,
using the data on the \br and \ber neutrino fluxes. 
Our results show, in particular, that 
the values of the parameters $x_{j_1}$ and $x_{j_2}$
thus determined could differ from their 
respective SSM values  by factors determined by 
the four logarithmic derivatives - 
$\alpha_{B,j_1}$, $\alpha_{B,j_2}$, $\alpha_{Be,j_1}$ and
$\alpha_{Be,j_2}$ (eqs. (\ref{xj1central_2}) and 
(\ref{xj2central_2})). 
We have calculated the SSM parameter uncertainties 
for all possible combinations
of two SSM parameters $\{x_{j_1}$,$x_{j_2}\}$,
assuming that $^8B$ and $^7Be$
neutrino fluxes are measured with 1$\sigma$ errors 
of  4\% and 6\%, respectively.
Such precision on $\phi_{Be}$ can be reached in 
the Borexino experiment.
We have found that the 
uncertainties of $x_{j_1}$ and $x_{j_2}$
are controlled in a rather complicated way
by both the magnitude and the 
relative signs of the different logarithmic
derivatives (eqs. (\ref{xj1error_2}) and (\ref{xj2error_2})). 
If for a certain pair of SSM parameters $x_{j1}$ and $x_{j2}$
the relation $\alpha_{B,j_2}/\alpha_{B,j_1}\cong
\alpha_{Be,j_2}/\alpha_{Be,j_1}$ holds, 
these parameters would be determined with 
poor accuracy even if one uses high 
precision data on the $^8B$ and $^7Be$ 
neutrino fluxes. The logarithmic derivatives
taken from the BP04 SSM (table \ref{tab_alpha}) 
suggest that such pairs can be, for instance,  
$\{S_{33},S_{34}\}$, $\{L_\odot,O_\odot\}$ 
and $\{Z/X,D_\odot\}$. 
We have found also that 
among the solar physics parameters 
the opacity $O_\odot$ can be determined with a
9\% uncertainty in pair with $S_{34}$, or 
$S_{17}$, or $S_{e^-7}$, for which we get
at the same time uncertainty of approximately 16\%.
For the diffusion $D_\odot$ we find an uncertainty of 11\%
when determined in combination with
$S_{33}$ or $S_{34}$. In these cases
the latter are found with uncertainties
of 36\% and 14\%, respectively. 
Most of the results from this part of our study are 
summarised in Table \ref{tab_combination}. 

  We have obtained also rather detailed
results on the uncertainties of the 
combination $\{Z/X$,$S_{34}\}$, determined 
from data on the $^8B$ and $^7Be$ neutrino fluxes
(Table \ref{tab_2param} and Fig. \ref{fig:combination}).
We have found, in particular, that
if the 1$\sigma$ uncertainties in
the values of $\phi_B$ and $\phi_{Be}$ 
are smaller than 4\%,
the uncertainty in the value of $S_{34}$,  
determined using the neutrino flux measurements, 
would be smaller than its presently estimated
uncertainty of $\sim$ 9\% . 
The uncertainty in the determination of 
$Z/X$ would be smaller than its currently
SSM estimated one of 15\% 
if $\phi_{Be}$ is measured with an 
error not exceeding approximately 
$10\%$ at 1$\sigma$.

  Finally, we have analyzed the possibility to use
high precision prospective measurements of the 
\br, \ber and $pp$ solar neutrino 
fluxes to simultaneously determine 
any three of the SSM input parameters.
We have taken different combinations of three SSM 
parameters and calculate their uncertainties using 
3\%, 4\% and 1\% as illustrative 1$\sigma$ errors in the 
measured values of $\phi_B$, $\phi_{Be}$ and $\phi_{pp}$,
respectively. We have found that for almost all 
solar model  parameters, 
the uncertainties reduce 
when using the combined information on the
$^7B$, $^8Be$ and $pp$ neutrino fluxes, compared to 
what we have obtained by  using 
only prospective data on $^8B$ and/or $^7Be$ fluxes. 
Results for those sets of three SSM input parameters 
which are determined with uncertainties smaller than
15\% each are collected in Table\ \ref{tab_combination3}.
Our results show, in particular, 
that the most precise determination of $S_{34}$ 
occurs in the combination $\{S_{34},L_\odot,S_{e^-7}\}$, 
while $Z/X$ is best determined in the set 
$\{S_{17},L_\odot,Z/X\}$. For both these parameters 
the uncertainties we get are smaller 
than those in the respective BP04 SSM predictions. 
The best determination of $S_{11}$ 
is found to be from the set 
$\{S_{11},,L_\odot,S_{e^-7}\}$, of $D_\odot$ from 
$\{S_{17},L_\odot,D_\odot\}$, and of $\tau_\odot$ from 
$\{S_{17},L_\odot,\tau_\odot\}$. 
Even though the precision we obtained for 
$L_\odot$ is worse than the precision achieved in the
direct measurement of $L_\odot$, 
the method used to determine $L_\odot$
allows to perform a
fundamental test of the thermo-nuclear fusion 
theory of energy generation in the Sun, 
as well as to test the hypothesis that the Sun
is in an approximate steady state 
in what regards the energy produced 
in its central region and the 
energy emitted from its surface.

  The results obtained in the present article 
underline the importance of  
performing high precision measurements of 
\ber and $pp$ solar neutrino fluxes.

\vspace*{5mm}{\bf Acknowledgments.}
We dedicate this paper to the memories of 
J.N. Bahcall and R. Davis.
S.C. and S.T.P. would like to thank S.M. Bilenky for useful discussions.
The collaboration of T. Schwetz at the initial stage of this
study is acknowledged with gratefulness.
This work was supported in part by the Italian MIUR and INFN under the
programs ``Fisica Astroparticellare'' (S.T.P.). The work of S.G. 
was supported by the Alexander--von--Humboldt--Foundation. 
The work of A.B. is partly supported by the
Partner Group program between the Max Planck Institute
for Physics and Tata Institute of Fundamental Research.

\end{document}